\documentclass{PoS}
\usepackage{graphicx}
\usepackage{amsmath}

\title{The background field method on the lattice}

\ShortTitle{Background field method}

\author{\speaker{Andrei Alexandru}\\
        The George Washington University, Washington, DC, USA\\
        E-mail: \email{aalexan@gwu.edu}}

\author{Frank X. Lee\\
        The George Washington University, Washington, DC, USA\\
        E-mail: \email{fxlee@gwu.edu}}

\abstract{The background field method 
has been used successfully to determine hadron electromagnetic polarizabilities. Recently questions have been
raised regarding the proper way to deal with the electric field on the lattice. In this paper, we show how the
presence of a background electric field affects the quark hopping matrix. We use this formulation to carry out
simulations on quenched configurations, and we present the results for neutron electric polarizability for pion 
masses as low as $500\MeV$. We find that the polarizability is roughly constant for the quark masses considered,
$\alpha\sim 1.5 \times 10^{-4} \fm^3$. While the polarizability is positive, it is significantly smaller than the 
experimental value due to the fact that the quark masses used are too large.}

\FullConference{The XXVI International Symposium on Lattice Field Theory\\
		 July 14-19 2008\\
		 Williamsburg, Virginia, USA}

\def\fm {\mathop{\hbox{fm}}}
\def\MeV {\mathop{\hbox{MeV}}}

\def\dd  {\mbox{d}}
\def\ra {\rightarrow}
\def\av  #1 {\left< #1 \right>}

\newcommand{\beq}{\begin{equation}}
\newcommand{\eeq}{\end{equation}}
\newcommand{\beqa}{\begin{eqnarray}}
\newcommand{\eeqa}{\end{eqnarray}}

\begin{document}

\section{Introduction}

Background field method has shown promising results in computing hadron electromagnetic moments and 
polarizabilities \cite{hrf1989,jch2005, fxl2005, fxl2006, men2007, detmold08, aubin08}. The basic idea \cite{hrf1989} is to introduce a static 
electromagnetic field on the lattice and then measure the change in the hadron masses due to the presence 
of the field. To the lowest orders, the shift in the hadron masses is given by
\beq
\Delta E = -\vec{p}\cdot\vec{E}-\vec{\mu}\cdot\vec{B}-\frac{1}{2}\left(\alpha E^2+\beta B^2\right) + ...,
\eeq  
where $\vec{p}$ and $\vec{\mu}$ are the electric and magnetic dipole moments and $\alpha$ and $\beta$ are 
the electric and magnetic polarizabilities. To introduce the field, the charged particles get coupled minimally with the
electromagnetic field; the covariant derivative becomes
\beq
D_\mu = \partial_\mu - i g G_\mu -i q A_\mu,
\eeq
where $G_\mu$ is the chromoelectric field and $A_\mu$ is the static electromagnetic background. On the lattice, the
presence of the electromagnetic background modifies the fermion hopping matrix; from a practical point of view, the
change can be viewed as an additional phase factor multiplying the links:
\beq
U_\mu\rightarrow e^{-i q a A_\mu}U_\mu.
\eeq
A complication arises due to the Euclidean nature of lattice formulation; {\em formally}, one has the following rules for
converting the Minkowski formulation to Euclidean:
\beqa
\nonumber
x_{1,2,3} \rightarrow x_{1,2,3} &\quad& A_{1,2,3} \rightarrow A_{1,2,3} \\
x_0 \rightarrow x_4 = i x_0 &\quad& A_0\rightarrow A_4 = -i A_0.
\eeqa
To introduce a constant electric field in the x-direction we can choose the following potential:
\beqa
\label{eq:1.5}
A_{M} = \left(0, +E t, 0, 0\right) &\rightarrow& A_E = \left(-i E x_4, 0, 0, 0\right) \quad {\rm or} \\
\label{eq:1.6}
A_{M} = \left(-E x, 0, 0, 0\right) &\rightarrow& A_E = \left(0, 0, 0, +i E x_1\right),
\eeqa
where $A_M$ stands for Minkowski and $A_E$ for Euclidean formulation; note that in the Minkowski 
formulation the components are $(A_0, \vec{A})$ and in the Euclidean one they are $(\vec{A}, A_4)$.
Similarly,  for a magnetic field in the x-direction we can use the following potential choices:
\beqa
A_{M} = \left(0, 0, +B z, 0\right) &\rightarrow& A_E=\left(0, +Bz, 0, 0\right) \quad {\rm or} \\
A_{M} = \left(0, 0, 0, -B y\right) &\rightarrow& A_E =\left(0, 0, -B y, 0\right).
\eeqa
Thus, on the lattice we can use the following phase factors to produce a constant electric or magnetic
field:
\beqa
\label{eq:1.9}
&E_x:& \quad U_1 \ra e^{-q a E x_4} U_1 \quad {\rm or} \quad U_4 \ra e^{q a E x_1} U_4 \\
&B_x:& \quad U_2 \ra e^{-iq a B x_3} U_2 \quad {\rm or} \quad U_3 \ra e^{i q a B x_2} U_3.
\eeqa
The surprising result in Eq.~(\ref{eq:1.9}) is that the phase factors are real, in contrast to the magnetic case
where the phase factors are $U(1)$ phases. This can be traced back to the extra $i$ factor that appear
when performing the rotation to the ``imaginary" time: for the first choice of potential it arises from the fact
that the time becomes imaginary, whereas for the second choice it comes from changing $A_0$ to $-i A_4$.
This result is at variance with the standard treatment 
\cite{hrf1989, jch2005, fxl2005, fxl2006, men2007} where both the electric and magnetic field are introduced 
using a $U(1)$ phase. The fact that the electric field defined using a $U(1)$ phase corresponds to an imaginary
electric field in Minkowski space was first pointed out by E. Shintani et all \cite{esh2007}.

The purpose of this paper is to show that the result presented above, derived using {\em formal} rules, is correct. 
In section \ref{sec2} we first argue for its correctness using the Wilson loop as a simple example and then carry 
out an explicit calculation for charged bosons placed in a constant electric field. 

In section \ref{sec3} we address the issue of the exponential versus linear phase factor. It is argued \cite{hrf1989} 
that the phase factors in Eq.~(\ref{eq:1.9}) should be replaced with a linearized version. We show 
that it is better to keep them in the exponential form.

Finally, in section \ref{sec4} we present the results of our simulations for the electric polarizabilities of the neutron. 
We show how to compute its polarizability using either the real phase factor or the $U(1)$ phase. We also present
the result for the linear case and compare it with the exponential case.

\section{Euclidean formulation\label{sec2}}

\begin{figure}[ht]
   \centering
   \includegraphics[width=4in]{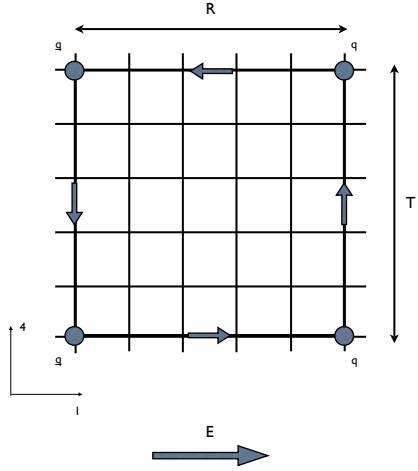} 
   \caption{Heavy quark--anti-quark pair in the presence of an electric field.}
   \label{fig:1}
\end{figure}

To show that the use of a real phase factor in the electric field case is justified imagine that we place a heavy
quark--anti-quark pair in a electric field (see Fig. \ref{fig:1}). For large $T$ the Wilson loop average
decays exponentially, i.e. $\av{W} \sim e^{-V(R) T}$. In the presence of an electric field the energy of the
state is altered by the dipole energy: $V(R)\ra V(R)-qER$, where $q$ is the charge of the quark. This defines our expectations. 

When the electric field field is introduced on the lattice using the first choice in Eq.~(\ref{eq:1.9}) the links in the
$x$ direction get an addition phase factor; the Wilson loop is then changed: the bottom links in Fig. \ref{fig:1} 
contribute a factor $(e^{-qaEt})^{R/a}=e^{-q E R t}$ and the top links contribute $(e^{qaE(t+T)})^{R/a}=e^{q E R(t+T)}$.
Since this factor is the same for every configuration, the average value of the Wilson loop becomes
\beq
\left<W\right>\rightarrow \left<W\right> e^{-q E R t} e^{q E R(t+T)}\sim
e^{-V(R)T+q E R T}.
\eeq
We see then that the potential is modified according to our expectations; the change in the energy due to the
presence of the electric field $\Delta m = - q E R$ is exactly the dipole energy. 

To further confirm that this is the correct way to introduce the electric field we can compute the partition function
in the Euclidean formalism for a charged bosonic field. The Lagrangian of the system is
\beq
L = \frac{1}{2}\int d^3x \left[\partial_\mu\phi^* \partial^\mu\phi - m^2 \phi^*\phi\right].
\eeq
We introduce the background field using the minimal substitution $\partial_\mu \ra \partial_\mu + i q A_\mu$.
The Lagrangian in the presence of the field becomes:
\beq
L = \int d^3x \left[(\partial_\mu-iq A_\mu)\phi^* (\partial^\mu+iq A^\mu)\phi - m^2 \phi^*\phi\right] ,
\eeq
where we see that the covariant derivative acting on $\phi^*$ is $\partial_\mu-iq A_\mu$ due to the fact that
the conjugate field has opposite charge (it is also needed to preserve the reality of the Lagrangian).

To get to the Euclidean formulation we need the Hamiltonian of the system which we get using the Legendre
transform 
\beq
H= \int \dd^3x\, \left[\pi(x) \dot{\phi}(x)+\pi^*(x) \dot{\phi}^*(x)\right] - L, 
\eeq
where $\pi(x)=\frac{\partial L}{\partial \dot{\phi}(x)}=\dot{\phi}^*(x)-iqA_0(x)\phi^*(x)$
\footnote{Note that while this is the standard representation for the charged Klein-Gordon field, the use
of complex variables is formal; the discussion is correctly carried out in terms of two real fields $\phi_{1,2}$ which
are related to our field by $\phi=(\phi_1+i\phi_2)/\sqrt{2}$. }. The Hamiltonian of this system is
\beq
H=\int \dd^3x \left[ \pi^*\pi + i q A_0 (\pi^*\phi^*-\pi\phi)+(\nabla - i q \vec{A})\phi^* (\nabla + iq \vec{A})\phi + 
m^2 \phi^*\phi\right]. 
\eeq
To quantify this system we discretize it, introduce field operators defined on lattice points satisfying the canonical
commutation relations and then normal order it. Since we are interested in the electric field we will set $\vec{A}=0$.
The resulting hamiltonian is:
\beq
\hat{H}=\sum_n \left[\hat\pi_n^*\hat\pi_n + i q A_0 (\hat\pi_n^*\hat\phi_n^*-\hat\pi_n\hat\phi_n)+(\tilde{\nabla}\hat\phi_n)^*\tilde{\nabla}\hat\phi_n+m^2 \hat\phi_n^*\hat\phi_n\right],
\eeq
where $[\hat{\phi}_n,\hat{\pi}_m]=i\delta_{n,m}$ and $\tilde{\nabla}$ is some discretization of the gradient.

To determine the Euclidean action we need to compute the matrix element
\beq
\av{\phi_{t+1} \left| e^{-a_t \hat{H}}\right| \phi_t} =e^{-a_t \sum_n {\cal L}_E(n,t)},
\eeq
where $\left| \phi\right>$ is a vector in the "field" representation 
(i.e. $\hat{\phi}_n \left|\phi\right> = \phi_n \left|\phi\right>$ and $\hat{\phi}_n^* \left|\phi\right> = \phi_n^* \left|\phi\right>$)
and ${\cal L}_E$ is the Euclidean Lagrangian density. The calculation is straightforward; the most difficult
steps are 
\beqa
&&\av{\phi' \left| e^{-a_t \sum_n \hat{\pi}_n^*\hat{\pi}_n}\right| \phi} =\prod_n \frac{1}{2\pi a_t} e^{-\frac{1}{a_t} |\phi'-\phi|^2}
\quad and \\
&&e^{-a_t i q A_0 (\hat{\pi}_n^*\hat{\phi}_n^*-\hat{\pi}_n\hat{\phi}_n)}\left| \phi\right> = \left| \tilde{\phi} \right>,
\eeqa
where $\tilde{\phi}_m = \phi_m$ for $m\not=n$, $\tilde{\phi}_n=e^{a_t q A_0} \phi_n$ and 
$\tilde{\phi}^*_n=e^{-a_t q A_0}\phi^*_n$
\footnote{Note that $\phi^*$ is not the complex conjugate of $\phi$; this confusion is due to our improper use of 
complex variables. We treat $\phi$ and $\phi^*$ as independent variables.}. 
Using the relations above we get
\beq
{\cal L}_E(n,t)=(\tilde{\nabla}\phi_{t,n})^*\tilde{\nabla}\phi_{t,n}+m^2\phi_{t,n}^*\phi_{t,n}+
\frac{1}{a_t^2} \left(\phi_{t+1,n}-e^{a_t q A_0}\phi_{t,n}\right)\left(\phi_{t+1,n}^*-e^{-a_t q A_0}\phi_{t,n}^*\right) + o(a_t^2).
\eeq
The important thing to notice is the electric field contribution appearing in the hopping term in the time direction.
There are no links in this formulation, but we see that the field propagating forward in time gets a factor due to 
the electric field $e^{a_t q A_0}$; this is exactly the same factor as the one derived using the formal rules for the
second choice of potential in Eq.~(\ref{eq:1.9}).

We have shown that in the case of charged bosons the phase factor due to an external electric field is real. 
We see no reason why this should be different for fermions so we conclude that the rules derived formally in the 
first section are correct. While we didn't show it,  the hopping matrix for a charge boson in the presence of a
magnetic field has a $U(1)$ factor; this is also in perfect agreement with the factors derived using the formal
rules.

\section{Exponential vs. linear phase factor \label{sec3}}

As mentioned in the Introduction, most of the studies carried out this far \cite{hrf1989, jch2005, fxl2005, fxl2006} use
a linearized form of the phase factor. It was suggested in \cite{hrf1989} that the exponential factor $e^{-iqaA_\mu}$ 
should be replaced by its linearized version $1-iqaA_\mu$; the authors argued that the the linearized factor 
makes the Dirac matrix resemble the continuum like covariant derivative where the coupling with the 
electromagnetic field is linear. 

The difference between the linear and exponential form is of the order $a^2$ so it seems like that shouldn't make
much of a difference in the continuum limit. However, the change to linear form also changes the action
at the order $E^2$; the polarizability is derived from the response of the observables at the order $E^2$ thus 
its value is affected. In the next section we will show that, at least when $a$ is around $0.1\fm$, this change 
can be quite significant.

It may be that in the continuum limit the polarizability is the same irrespective of how we introduce the electric field.
However, this is not guaranteed; for example, when computing the energy density of a free fermionic gas at non-zero
chemical potential the lattice result is wrong when the chemical potential is introduced using an expression similar
to the one in the continuum \cite{Hasenfratz:1983ba}. While this example might look off-topic, it is in fact very pertinent
to our discussion: the chemical potential plays a very similar role to $A_0$. Hasenfratz and Karsch
\cite{Hasenfratz:1983ba} point out this 
similarity and argue that it is the gauge symmetry in $A_0$ that protects the continuum like formulation from 
divergencies. They conclude that in order to avoid divergencies on the lattice, you have to couple the chemical
potential in a gauge invariant way: the phase factors are then changed from $1\pm a\mu$ to $e^{\pm a\mu}$.

While it is not entirely clear that their argument can be carried over to our situation, it is worth pointing out that when
using the linearized version the gauge symmetry in the external field is lost. To be more precise, take the two 
choices for the electromagnetic potential in Eqs.~(\ref{eq:1.5}) and (\ref{eq:1.6}); they both represent an electric field
in the x-direction and we expect that the results that they produce are the same. If we are to use the corresponding 
exponential factors in Eq.~(\ref{eq:1.9}) then the results come out to be the same since the gauge symmetry is preserved.
If we choose to linearize these factors then the results of our simulations would be different. 
The Wilson loop in the presence of the electric field introduced using
the linear factors is:
\beqa
U_1:\quad\left< W\right> &\sim& e^{-V(R)T+q E R T} e^{\frac{1}{2}a q^2 E^2 R\left[(t+T)^2+t^2\right]} \\
U_4:\quad\left< W\right> &\sim& e^{-V(R)T+q E R T} e^{\frac{1}{2}a q^2 E^2 T\left[(x+R)^2+x^2\right]}
\eeqa 
depending on whether we introduced it using the space links $U_1$ or the time links $U_4$. Notice that when we
use the electric field using time links the Wilson loop decays exponentially in time and the only difference is an $a$
dependent contribution to the dipole energy; more worrisome is the fact that when using the space links the
Wilson loop doesn't decay exponentially with time but it acquires a piece that goes like $e^{-T^2}$. This would make
it impossible to determine the energy of the state by simply fitting for an exponential behavior.

We conclude that it is better to use the exponential form rather than the linear form. Even if the continuum limit is
the same for both formulations (which is not guaranteed), the exponential form has the benefit of being 
gauge invariant in the external field.

\section{Numerical results \label{sec4}}

To check these ideas we ran a set of quenched simulations on $24^4$ lattices. We used Wilson action at
$\beta=6.0$ which corresponds to a lattice spacing of $a=0.093\fm$. We used clover fermions and our lowest 
pion mass was $500\MeV$. For the electric field we used $\eta=a^2 q E = 0.00576$, where $q$ is the 
charge of the down quark. To avoid dealing with field discontinuity we used Dirichlet boundary conditions in
both time and x direction.

We computed the correlation function for the neutron $G_E(t)$ in the presence of the electric field $E$. 
To extract the polarizability we computed the ratio $G_E(t)/G_0(t)$ which is expected to behave asymptotically
as $e^{-\Delta m\, t}$. Unfortunately, our lattice was not long enough in the time direction and we used a two 
exponential form to fit $G_E$ and $G_0$.

\begin{figure}[h]
   \centering
   \includegraphics{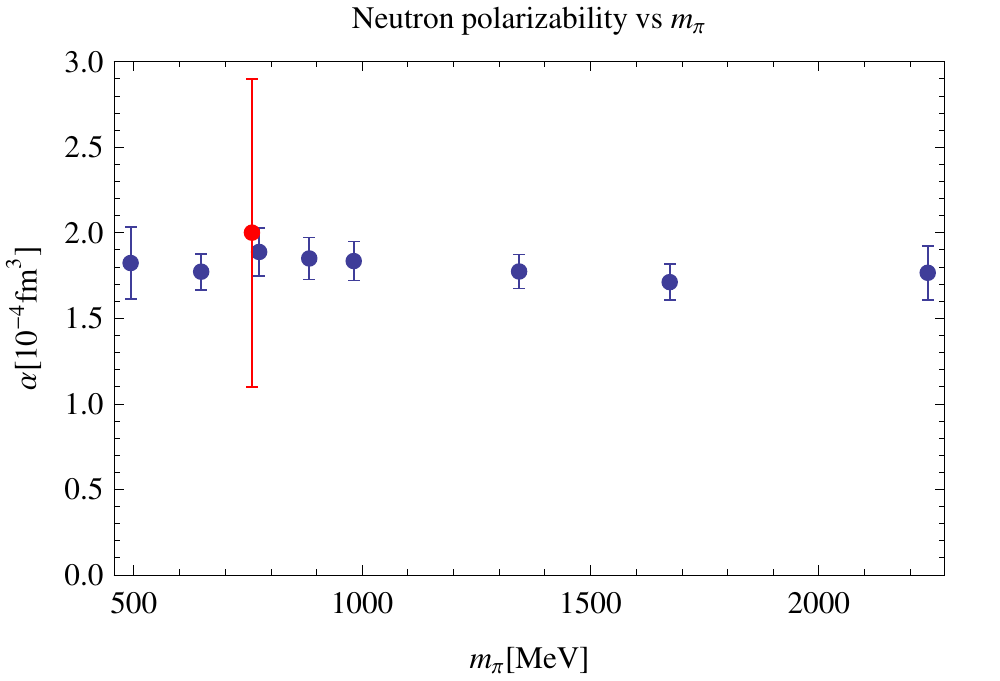} 
   \caption{Neutron polarizability as a function of the pion mass. The result in red
   includes the dynamical effects of the fermions and the influence of the electric field on the vacuum \cite{men2007}.}
   \label{fig:2}
\end{figure}

After extracting the mass shift we computed the polarizability. In Fig. \ref{fig:2} we plot the polarizability as a 
function of the pion mass. Note that the polarizability is roughly constant in the range of quark masses we 
studied. Also plotted is a result that includes dynamical fermion effects and the effect of fermions on the background
\cite{men2007}
\footnote{The value reported in \cite{men2007} is the negative of the value plotted here -- this is due to the 
confusion regarding the proper way of introducing the electric field in Euclidean formulation.}. We see that our
result is in good agreement with this value and that the dynamical effects do not play an important role when
the pion mass is around $650\MeV$; this is going to change when we approach the chiral limit.

We also run a set of simulations using the same parameters but using a $U(1)$ factor to introduce the electric
field -- this corresponds to introducing a imaginary electric field $iE$. The expectation was that the mass shift is 
going to change sign since it depends on the square of the electric field. We verified that this is indeed true and
the polarization derived using these mass shift agreed perfectly with the one derived using the real phase factor.

\begin{figure}[h]
   \centering
   \includegraphics{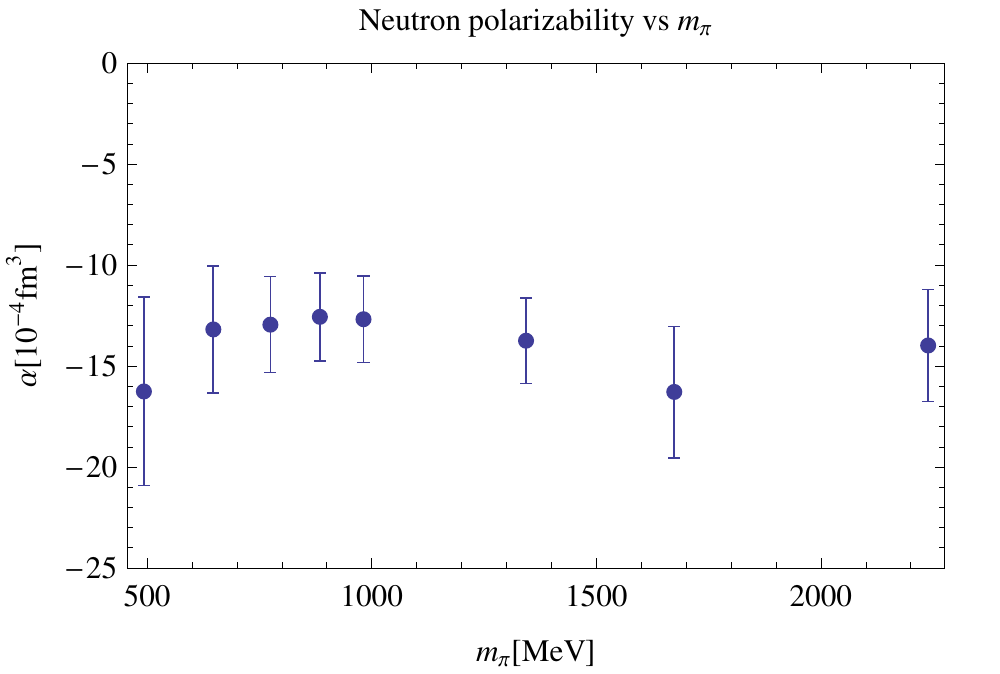} 
   \caption{Neutron polarizability as a function of the pion mass when introducing the electric field via a linear factor.}
   \label{fig:3}
\end{figure}

The final set of simulations used a linear expression for the electric field contribution. We argued in the previous 
section against using it, but we wanted to check whether the difference is significant. In Fig. \ref{fig:3} we plot the
polarizability computed with the linear field. We first note that these values agree well with the results reported in
\cite{jch2005} except for the sign; the sign difference arises from the confusion regarding the electric field in Euclidean
time. We see that these results no longer look like they are converging towards the experimental value -- even the sign
differs. We also see that the difference between the linear and exponential case is quite significant; in the unlikely case
that they converge to the same value it would take extremely fine lattices given the discrepancy for $a=0.093\fm$.

Our numerical simulations support the conclusion of our previous section: to introduce the electric
field we need to use an exponential phase factor; the linear case seems to differ substantially from the exponential case. It is also
important to note that we can compute the polarizability using a $U(1)$ factor as long as we remember that this
corresponds to an imaginary electric field.

\section{Conclusion and outlook}

Our goal was to understand how to relate the electric field in the Minkowski space -- the physical electric field -- to the
phase factors that appear in the hopping matrix. We have shown that the presence of an external electric field gives
rise to an exponential phase factor $e^{\pm aqEt}$. To compute the polarizability we can use either the real phase 
factors or $U(1)$ ones as long as we correct for the sign:
\beqa
U_1 \rightarrow U_1 e^{-aqEt}\Rightarrow \Delta m = -\frac{1}{2} \alpha E^2, \\
U_1 \rightarrow U_1 e^{-iaqEt}\Rightarrow \Delta m = +\frac{1}{2} \alpha E^2.
\eeqa
The presence of the magnetic field gives rise to a $U(1)$ factor and to compute the magnetic polarizability we use
\beq
U_2 \rightarrow U_2 e^{-i a q B x_3}\Rightarrow \Delta m = -\frac{1}{2} \beta B^2.
\eeq

Our numerical results agree well with previous simulations if we account for the sign change. The nucleon polarizability 
for $m_\pi = 500\MeV$ is positive but significantly smaller than the experimental value. This is actually in agreement 
with chiral perturbation theory that predicts a $1/m_\pi$ raise in polarizability as one approaches the chiral limit. We
don't see the trend yet and most likely we need to carry out simulations at smaller quark masses. The other major 
challenge is to include the contribution of the electric field on the vacuum which we expect to have an even bigger 
impact on the value of the polarizability than the inclusion of dynamical fermions. Our current plan is to include this
effect using a re-weighting of the determinant.

\end{document}